\documentclass[conference]{IEEEtran}
\IEEEoverridecommandlockouts

\usepackage[ruled,vlined]{algorithm2e}
\usepackage{array}
\usepackage{booktabs}
\usepackage{caption}
\usepackage{cite}
\usepackage{amsmath,amssymb,amsfonts}
\usepackage{hyperref}
\usepackage{graphicx,epstopdf}
\usepackage{multirow}
\usepackage{newtxmath}
\usepackage{subcaption}
\usepackage{textcomp}
\usepackage{xcolor}
\def\BibTeX{{\rm B\kern-.05em{\sc i\kern-.025em b}\kern-.08em
    T\kern-.1667em\lower.7ex\hbox{E}\kern-.125emX}}

\graphicspath{ {figures/} }

\newcommand{\update}[1]{\color{black}{#1}\color{black}}
\newcommand{\eg}{\emph{e.g.},\xspace}

\newcommand{\myparagraph}[1]{\vspace{0.5em}\noindent\textbf{#1}}
\newcommand{\ie}{\emph{i.e.,}\xspace}

\newcommand{\etal}{\emph{et~al.}\xspace}
\newcolumntype{x}[1]{>{\raggedright\arraybackslash}p{#1}}

\SetCommentSty{mycommfont}

\makeatletter 
\g@addto@macro{\@algocf@init}{\SetKwInOut{Output}{Output}} 
\makeatother    

\begin{document}

\title{\update{Multi-spatial Scale Event Detection from Geo-tagged Tweet Streams via Power-law Verification}}

\author{\IEEEauthorblockN{Yi Han, Shanika Karunasekera, Christopher Leckie, Aaron Harwood}
\IEEEauthorblockA{\textit{School of Computing and Information Systems} \\
\textit{The University of Melbourne}\\
Parkville, Australia \\
\{yi.han, karus, caleckie, aharwood\}@unimelb.edu.au}
}

\maketitle

\begin{abstract}
Compared with traditional news media, social media nowadays provides a richer and more timely source of news. We are interested in multi-spatial level event detection from geo-tagged tweet streams. Specifically, in this paper we (1) examine the statistical characteristic for the time series of the number of geo-tagged tweets posted from specific regions during a short time interval, \eg ten seconds or one minute; (2) verify from over thirty datasets that while almost all such time series exhibit self-similarity, those that correspond to events, especially short-term and unplanned outbursts, follow a power-law distribution; (3) \update{demonstrate that these findings can be applied to facilitate event detection from tweet streams. We propose two algorithms---\textit{Power-law basic} and \textit{Power-law advanced}, where \textit{Power-law basic} only checks the existence of power-law distributions in the time series from tweet streams at multi-spatial scales, without looking into the content of each tweet, and \textit{Power-law advanced} integrates power-law verification with semantic analysis via word embedding. Our experiments on multiple datasets show that by considering spatio-temporal statistical distributions of tweets alone, the seemingly naive algorithm of \textit{Power-law basic} achieves comparable results with more advanced event detection methods, while the semantic analysis enhanced version, \textit{Power-law advanced}, can significantly increase both the precision and the recall.}
\end{abstract}


\section{Introduction}
Social media, especially Twitter, has become an increasingly more popular source of news. Compared with traditional forms of media, such as TV and newspapers, it often provides more timely information about various type of incidents. We are interested in real-life event detection at multi-spatial levels from geo-tagged tweet streams. 

We initially investigated using Poisson models to monitor the fluctuations in the time series of the number of geo-tagged tweets posted within a bounding box during a short time interval, \eg ten seconds to one minute. However, our experimental results observe a relatively high false positive rate for this Poisson model based event detection method. This observation motivates us to reexamine the properties of these time series. Specifically, this paper aims to answer the following questions:

\myparagraph{Section~\ref{sec:self_similarity}: What are the statistical characteristics of the time series?} A draw of several time series at different time scales, \ie the number of tweets posted every 1, 10, 60, 1000 seconds, shows that burstiness persists over all these scales, which indicates self-similarity \cite{crovella_explaining_1995, willinger_self-similarity_1995}. In order to verify this finding, we collect 33 tweet datasets of different types, generate the corresponding time series by counting the number of tweets posted every minute, and check self-similarity using three popular methods~\cite{karagiannis_selfis:_2002,karagiannis_user-friendly_2003}: aggregate variance, R/S and Whittle (please refer to Section~\ref{sec:background} for a more detailed description on self-similarity and the three methods). Our results suggest that all the time series are self-similar.

\myparagraph{Section~\ref{sec:power_law}: Can the time series be better characterised by other models than the Poisson process?} The existence of self-similarity suggests that Poisson models are inadequate to capture the underlying dynamics in tweet streams. Instead, we examine whether a power-law distribution can be validated from these time series, and find that when an event occurs, it is indeed more likely to observe a power-law distribution in the time series generated from geo-tagged tweet streams.

\myparagraph{Section~\ref{sec:application}: Can the answers to the previous two questions be applied for event detection from geo-tagged tweet streams?} \update{We propose two event detection methods---\textit{Power-law basic} and \textit{Power-law advanced}: (1) \textit{Power-law basic} only checks the existence of power-law distributions in the tweet stream at multi-spatial scales, without looking into the content of each tweet, or using any other information except the geo-location. Our experiments demonstrate that when combined with a Quad-tree~\cite{finkel_quad_1974,samet_quadtree_1984}, this seemingly naive approach can achieve comparable performance with Geoburst~\cite{zhang_geoburst:_2016}\footnote{Although the improved versions exist (Geoburst+~\cite{zhang_geoburst+:_2018}, TrioVec~\cite{zhang_triovecevent:_2017}), we do not use them as baselines in this work as they are supervised approaches.}, a widely cited event detection algorithm that considers temporal, spatial and semantic information; (2) \textit{Power-law advanced} improves the algorithm by incorporating semantic analysis via word embedding, and our results suggest that it can significantly increase both the precision and the recall.}

The remainder of this paper is organised as follows: Section~\ref{sec:background} provides background information on self-similarity and power-law distributions; Section~\ref{sec:self_similarity} describes the collected datasets, and checks whether the generated time series exhibit self-similarity; Section~\ref{sec:power_law} verifies the power-law hypothesis; Section~\ref{sec:application} proposes two multi-spatial event detection algorithms, \textit{Power-law basic} and \textit{Power-law advanced}; Section~\ref{sec:related_work} summarises related work in event detection from social media; and Section~\ref{sec:conclusions} concludes the paper and gives directions for future work.

\section{Background on Self-similarity \& Power-law Distributions}\label{sec:background}
In this section, we briefly introduce the fundamental concepts in self-similarity and power-law distributions, including their definitions and the methods to verify them.

\subsection{Self-similarity}
Unlike traditional Poisson traffic, where short-term fluctuations average out over a longer period of time, self-similar traffic maintains burstiness at all time scales.

\subsubsection{Definition}
Before giving the definition of self-similarity, we first introduce the concept of an aggregated process: given a process \(X = (X_{i}: i = 1, 2, 3, ...)\), its aggregated process is \(X^{(m)} = (X_{k}^{(m)}, k = 1, 2, 3, ...)\), where \(m\) is the block size, \(X_{k}^{(m)} = \frac{1}{m} \sum_{j=m \cdot (k-1)+1}^{m\cdot k} X_{j} \). In other words, \(X^{(m)}\) partitions the original series \(X\) into non-overlapping segments of size \(m\), and then averages over each segment.

A process \(X\) is called \emph{exactly second-order self-similar}~\cite{crovella_explaining_1995,willinger_self-similarity_1995} with parameter \(H = 1 - \beta/2\), \(0 < \beta < 1\), if \(R_{m}(k) = R(k) \sim k^{-\beta}\) as \(k \rightarrow \infty\), where \(R_{m}(\cdot)\) and \(R(\cdot)\) are the autocorrelation functions for \(X^{(m)}\) and \(X\), respectively. The parameter \(H\) is called the Hurst parameter~\cite{hurst_long-term_1950,hurst_long-term_1965}, and for a self-similar process, \(H \in (0.5, 1)\).

\subsubsection{Methods to Test Self-similarity}
SELFIS~\cite{karagiannis_selfis:_2002,karagiannis_user-friendly_2003} is a popular tool for testing self-similarity. It provides a number of methods to calculate the Hurst parameter, and the following three widely used methods are selected in this paper.

\begin{itemize}
    \item Aggregate variance. A sufficient condition of self-similarity is \(V_{m} = V \cdot m^{-\beta},\ m \in Z^{>1} = \{2, 3, ...\}\), where \(V\) (\(V_{m}\)) is the variance of \(X\) (\(X^{(m)}\)). Therefore, if a log-log plot of \(V_{m}/V\) is drawn against \(m\), then a straight line with slope \(\beta\) larger than \(-1\) indicates self-similarity, and \(H = 1 - \beta/2\).
    \item \(R/S\). For a self-similar process, its \emph{rescaled adjusted range} or R/S statistic can be represented by the relation:
    \begin{center}
    \(\lim_{n\to\infty} E(\frac{R}{S}(n)) = C\cdot n^{H}\),
    \end{center}
    where \(C\) is a finite positive constant, and \(n\) is the number of points in the process. Therefore, in the log-log plot of \(\frac{R}{S}\) against \(n\), the slope is an estimate of \(H\).
    \item Whittle. The Whittle method applies maximum likelihood estimation to the spectral density function of \(X\). It not only estimates \(H\), but also produces a confidence interval.
\end{itemize}

\subsection{Power-law Distribution}
A power-law probability distribution~\cite{clauset_power-law_2009,virkar_power-law_2014} takes the form of \(p(x) \propto x^{-\alpha}\), where \(\alpha\) is a positive constant often known as the exponent or scaling parameter.

A straightforward way to visualise a power-law distribution is to draw a log-log plot of the complementary cumulative distribution function (CCDF), and a roughly straight line is expected to be seen in the plot. However, this is only a necessary but not sufficient condition for a power-law distribution.

In order to validate that a time series \(X\) follows a power-law distribution, we first fit a power-law model to \(X\), and then run the Kolmogorov-Smirnov (KS) test~\cite{press_numerical_2007}. If the \(p\)-value of this significance test is below 0.05, the power-law hypothesis is rejected. Note that if a time series follows a power-law distribution, its log-log transformed CCDF is expected to be qualitatively similar at different scales, and hence it also exhibits self-similarity. 

\section{Verification of Self-similarity in Tweet Streams}\label{sec:self_similarity}
In order to study the statistical characteristic of tweet streams, we have collected the following datasets:

\begin{itemize}
    \item \(D_{1}-D_{30}\) (public)---A collection of 30 datasets associated with real-world events from 2012 to 2016, each of which contains from around \(2 \times 10^{5}\) to nearly \(3 \times 10^{7}\) tweets~\cite{zubiaga_longitudinal_2018}.
    \item \(D_{31}\) (public)---Twitter event detection dataset, which contains more than 120 million tweets (although the majority of these tweets are not associated with any event). The ground truth for 506 events and associated tweets are given~\cite{mcminn_building_2013}.
    \item \(D_{32}\)---Twitter dataset shared by the authors of~\cite{zhang_geoburst+:_2018}, which includes 9.5 million geo-tagged tweets from New York between 2014-08-01 and 2014-11-30.
    \item \(D_{33}\)---Over 920 thousand geo-tagged tweets collected from Melbourne between 2014 and 2018.
\end{itemize}

These datasets cover different types of tweets: \(D_{1}-D_{30}\) contain tweets that are only associated with specific events, the majority of which have worldwide impact; \(D_{31}\) contains tweets that both do and do not correspond to (mostly) local or less influential events, with the ground truth provided; \(D_{32}\ \&\ D_{33}\) include all geo-tagged tweets from a region during a certain period of time. 

Note that because the original datasets of \(D_{1}-D_{31}\) only include tweet ids, we have used a tool called ``twitter-dataset-collector"~\cite{papadopoulos_symeon:_2019} to download all the tweets. Since these datasets consist of hundreds of millions of tweets, it is infeasible to use the Twitter API to collect them due to the rate limit---it takes a significant amount of time to retrieve all the tweets. Instead, the tool crawls the webpages and reconstructs the original tweets. However, some tweets have already been deleted, and hence are not retrievable in this way. Even for those obtained tweets, the collected information is not as rich as in what the API returns. For example, most of the tweets do not have any location information, and the rest only have a location label---normally a city/town name, rather than specific coordinates. In addition, the second is truncated in the publication time.

For each of the above datasets, we count the number of tweets posted every minute, generate the corresponding time series, and test whether they exhibit self-similarity using three methods: aggregate variance, R/S and Whittle.

\subsection{Self-similarity in \texorpdfstring{\(D_{1}-D_{30}\)}{Lg}}
\begin{figure}[ht!]
\centering

\begin{subfigure}{.9\columnwidth}
  \centering
  \includegraphics[width=\columnwidth]{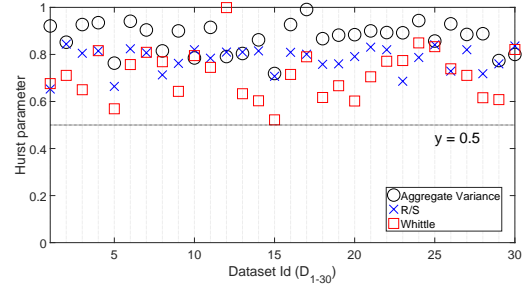}
  \caption{All tweets.}
  \label{figure_hurst_event_2012_2016_all}
\end{subfigure}

\begin{subfigure}{.9\columnwidth}
  \centering
  \includegraphics[width=\columnwidth]{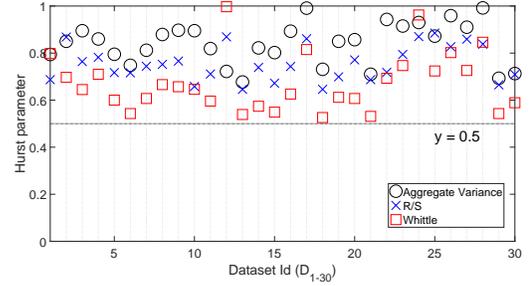}
  \caption{Tweets with a location label.}
  \label{figure_hurst_event_2012_2016_geo}
\end{subfigure}

\begin{subfigure}{.9\columnwidth}
  \centering
  \includegraphics[width=\columnwidth]{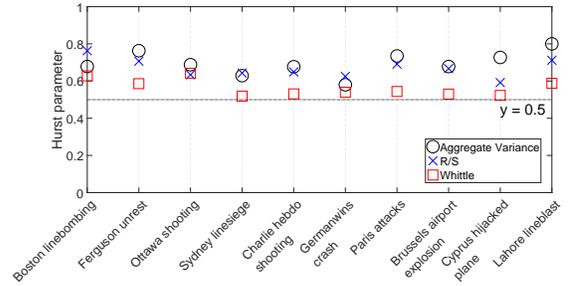}
  \caption{Local tweets (tweets close to where the event occurred), \ie for the events of (1) Boston marathon bombing, (2) Ferguson unrest, (4) Ottawa shooting, (5) Sydney siege, (6) Charlie Hebdo shooting, (7) Germanwings crash, (8) Paris attacks, (9) Brussels airport explosion, (10) Cyprus hijacked plane and (11) Lahore blast, only the tweets from Massachusetts, Ferguson, Ottawa, Sydney, Paris, France, Paris, Brussels, Cyprus, Lahore or Punjab are counted.}
  \label{figure_hurst_event_2012_2016_local}
\end{subfigure}

\caption{Hurst parameter estimates for \texorpdfstring{\(D_{1}-D_{30}\)}{Lg}. Events: 1. Boston marathon bombing; 2. Ferguson unrest; 3. Gaza under attack; 4. Ottawa shooting; 5. Sydney siege; 6. Charlie Hebdo shooting; 7. Germanwings crash; 8. Paris attacks; 9. Brussels airport explosion; 10. Cyprus hijacked plane; 11. Lahore blast; 12. Euro 2012; 13. Ebola outbreak; 14. Hong Kong protests; 15. Refugee Welcome; 16. Panama papers; 17. Hurricane Sandy; 18. Typhoon Hagupit; 19. Hurricane Patricia; 20. Nepal Earthquake; 21. Sismo Ecuador; 22. Mexican election 2012; 23. Obama and Romney 2012; 24. Superbowl 2012; 25. SXSW 2012; 26. US election 2012; 27. Indyref 2014; 28. St. Patrick's Day 2014; 29. Brexit; 30. Irish election 2016.}
\label{figure_hurst_event_2012_2016}
\end{figure}

We start with the 30 datasets of real-world events with wide impact. As can be seen from Fig.~\ref{figure_hurst_event_2012_2016_all}, all the estimates of the Hurst parameter are within the range of 0.5 and 1, which indicates that the corresponding time series are self-similar. Since we are more interested in geo-tagged tweets, we further examine the tweets that have a location label. The results in Fig.~\ref{figure_hurst_event_2012_2016_geo} suggest that these time series are self-similar too.

Among the 30 events, eleven of them are relatively short-term (from a few days to a couple of weeks), unplanned outbursts: Boston marathon bombing, Ferguson unrest, Gaza under attack, Ottawa shooting, Sydney siege, Charlie Hebdo shooting, Germanwings crash, Paris attacks, Brussels airport explosion, Cyprus hijacked plane and Lahore blast. This is the type of event that we are mostly interested in detecting from tweet streams. Therefore, for these eleven datasets, we extract tweets from close to where the events occurred, as those tweets will be most helpful in event detection. The results in Fig.~\ref{figure_hurst_event_2012_2016_local} show that the corresponding time series exhibit self-similarity as well. Note that for the event of ``Gaza under attack", insufficient data are collected locally, and hence it is not included.

\subsection{Self-similarity in \texorpdfstring{\(D_{31}\)}{Lg}}
\begin{figure}[ht!]
\centering
\includegraphics[width=.9\columnwidth]{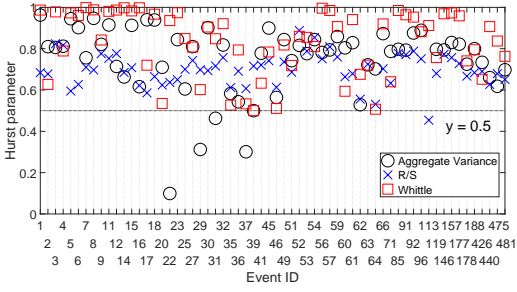}
\caption{Hurst parameter estimates for \texorpdfstring{\(D_{31}\)}{Lg}. Out of the 506 events, 62 of them meet all the three requirements: (1) lasting for a minimum of 64 minutes, (2) having at least 50 tweets, and (3) the maximum value in the time series is not smaller than 10. For a detailed description of the events, please refer~\cite{mcminn_building_2013}.}
\label{figure_hurst_event_2012}
\end{figure}

Next we examine the 506 events in the dataset of \(D_{31}\), the majority of which are local or less influential compared to those in \(D_{1}-D_{30}\). The tool SELFIS requires that a time series should have a minimum length of 64---in our case, since we count the number of tweets posted every minute, this means the event needs to last for at least 64 minutes (suggested by the collected data). However, quite a number of events in \(D_{31}\) do not meet this requirement, and hence are not considered. In addition, we also remove events that have less than 50 tweets, or whose time series have a maximum value of less than 10---never did 10 or more tweets get posted about the event within one minute. Finally, 62 events satisfy all three requirements, and the estimates of the Hurst parameter for the corresponding time series are shown in Fig.~\ref{figure_hurst_event_2012}.

As can be seen from the figure, five out of the 186 estimates are below 0.5. We believe that these outliers can be due to a lack of data for the five events: three of the time series have only 64 data points, while the other two have 128. For all events the length of whose time series is equal to or larger than 256, the estimates are all within the rage of 0.5 and 1.

We also calculate the Hurst parameter for the time series of the whole dataset of \(D_{31}\), since the majority of the 120 million tweets are not associated with the 506 events, and the estimates are also between 0.5 and 1. 

\subsection{Self-similarity in \texorpdfstring{\(D_{32}\ \&\ D_{33}\)}{Lg}}
We further test self-similarity in the datasets of geo-tagged tweets collected from New York and Melbourne. The statistics in Table~\ref{table_melb_ny} suggest that the time series generated from these two datasets also show self-similarity. Specifically, we not only check the overall case (level 0), but also zoom into sub-regions by recursively dividing the area into four equal parts (levels 1 to 3). The results indicate that in the city and all sub-region levels, their corresponding time series are self-similar. Table~\ref{table_melb_ny} lists part of the statistics.

In summary, our results in this section demonstrate that \textbf{\emph{self-similarity widely exists in different types of Twitter datasets, in terms of the number of tweets posted every minute}}. This conclusion can be extended to different time intervals due to the self-similarity. For example, for \(D_{32}\ \&\ D_{33}\), since each tweet's exact publication time (to the precision of a second) is known, we also check the time series of the number of tweets posted every 10 and 100 seconds, and the results also indicate self-similarity.

\begin{table}[t!]
\caption[caption]{Hurst parameter estimates for \texorpdfstring{\(D_{32}-D_{33}\)}{Lg}.
Level 0: the whole area;
Level 1: dividing the whole city into four equal sub-regions: 1-1, 1-2, 1-3, 1-4;
Level 2: dividing sub-region 1-1 into four equal parts: 2-1, 2-2, 2-3, 2-4;
Level 3: dividing sub-region 2-1 into four equal parts: 3-1, 3-2, 3-3, 3-4.}
\begin{center}
\begin{tabular}{|c|c|c|c|c|c|c|}
\hline
\multirow{3}{*}{Level} & \multicolumn{3}{c|}{\textbf{\(D_{32}\): New York}}  & \multicolumn{3}{c|}{\textbf{\(D_{33}\): Melbourne}} \\
\cline{2-7} 
 & Aggregate & \multirow{2}{*}{R/S} & \multirow{2}{*}{Whittle} & Aggregate & \multirow{2}{*}{R/S} & \multirow{2}{*}{Whittle} \\
 & Variance & & & Variance & & \\
\hline
0 & 0.68 & 0.68 & 0.97 & 0.90 & 0.75 & 0.60 \\
\hline
1: 1-1 & 0.90 & 0.78 & 0.89 & 0.87 & 0.72 & 0.57 \\
\hline
1: 1-2 & 0.90 & 0.80 & 0.84 & 0.87 & 0.72 & 0.57 \\
\hline
1: 1-3 & 0.92 & 0.80 & 0.84 & 0.79 & 0.66 & 0.55 \\
\hline
1: 1-4 & 0.90 & 0.79 & 0.87 & 0.82 & 0.67 & 0.56 \\
\hline
2: 2-1 & 0.88 & 0.80 & 0.85 & 0.85 & 0.70 & 0.56 \\
\hline
2: 2-2 & 0.90 & 0.81 & 0.82 & 0.85 & 0.71 & 0.56 \\
\hline
2: 2-3 & 0.90 & 0.82 & 0.77 & 0.77 & 0.66 & 0.55 \\
\hline
2: 2-4 & 0.89 & 0.82 & 0.79 & 0.78 & 0.65 & 0.55 \\
\hline
3: 3-1 & 0.85 & 0.82 & 0.82 & 0.81 & 0.67 & 0.55 \\
\hline
3: 3-2 & 0.86 & 0.82 & 0.77 & 0.80 & 0.68 & 0.55 \\
\hline
3: 3-3 & 0.84 & 0.84 & 0.71 & 0.76 & 0.65 & 0.56 \\
\hline
3: 3-4 & 0.87 & 0.82 & 0.75 & 0.72 & 0.59 & 0.54 \\
\hline
\end{tabular}
\label{table_melb_ny}
\end{center}
\end{table}

\section{Existence of Power-law Distribution in Event Tweet Streams}\label{sec:power_law}
In this section, we first examine whether the generated time series from Twitter datasets follow a power-law distribution. If this is the case, it explains self-similarity---a time series that follows a power-law distribution is also self-similar. Second, we reveal an important finding that when an event occurs it is much more likely to observe a power-law distribution in the tweet stream, compared with when no event occurs. This finding suggests that the existence of a power-law distribution can be used to help event detection from tweet streams.

Recall that in order to test the power-law hypothesis, we follow the approach introduced in Section~\ref{sec:background}: run the significance test, calculate the \(p\)-value for the fitted power-law model, and reject the hypothesis if the \(p\)-value is smaller than 0.05.

\subsection{Power-law Distribution in \texorpdfstring{\(D_{1}-D_{30}\)}{Lg}}
We still start with the 30 datasets of real-world events (\(D_{1}-D_{30}\)). However, we only check the existence of power-law distribution for tweets with a location label, which are useful in local event detection.

\begin{figure}[t!]
\centering
\begin{subfigure}{.9\columnwidth}
  \centering
  \includegraphics[width=\columnwidth]{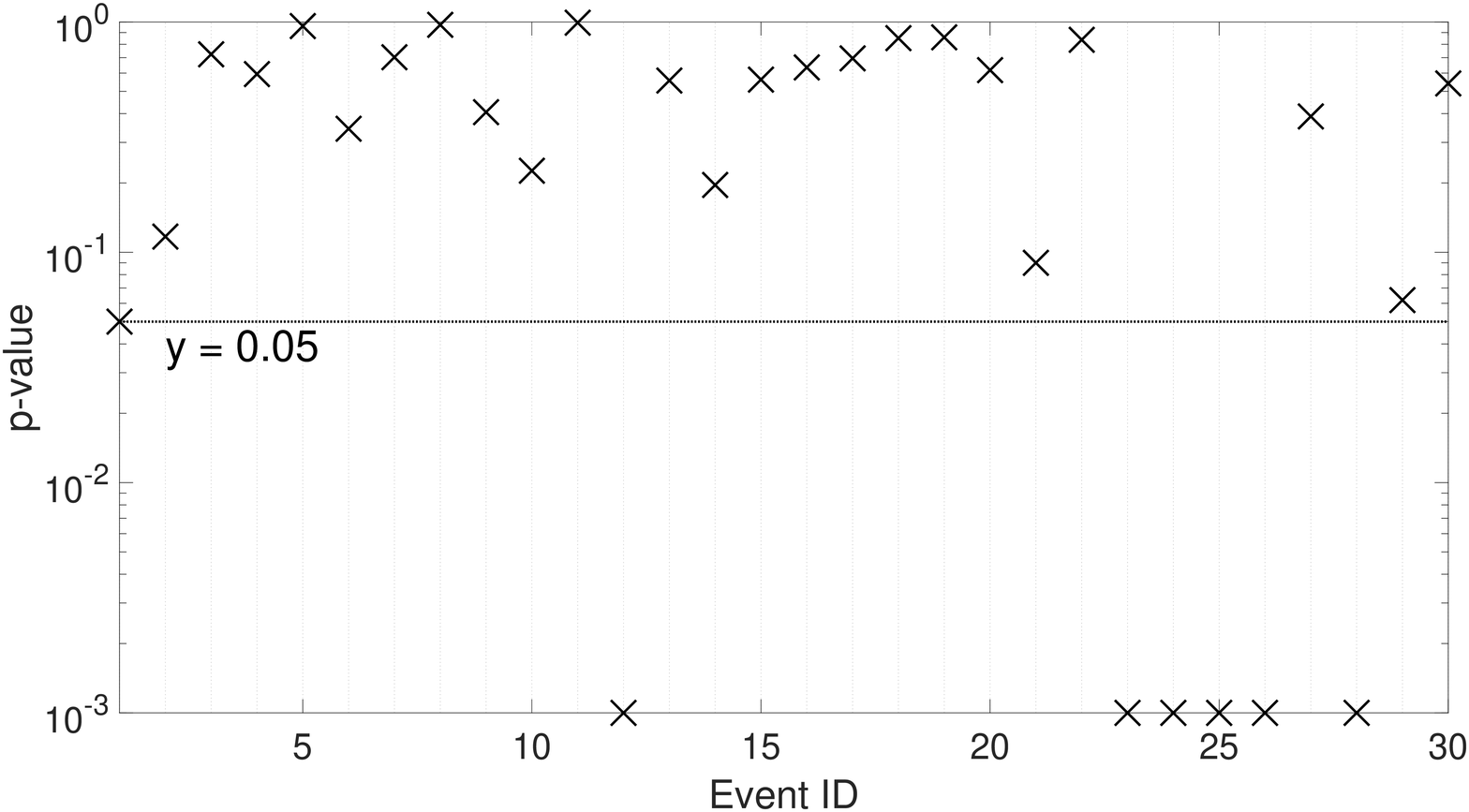}
  \caption{Tweets with a location label. For Events 12 (Euro 2012), 23 (Obama and Romney), 24 (Superbowl 2012), 25 (SXSW 2012), 26 (US election) and 28 (St. Patrick's Day 2014), the original \(p\)-values of 0.0 are replaced with 0.001, in order for them to be plotted.}
  \label{figure_p_value_2012_2016_geo}
\end{subfigure}

\begin{subfigure}{.9\columnwidth}
  \centering
  \includegraphics[width=\columnwidth]{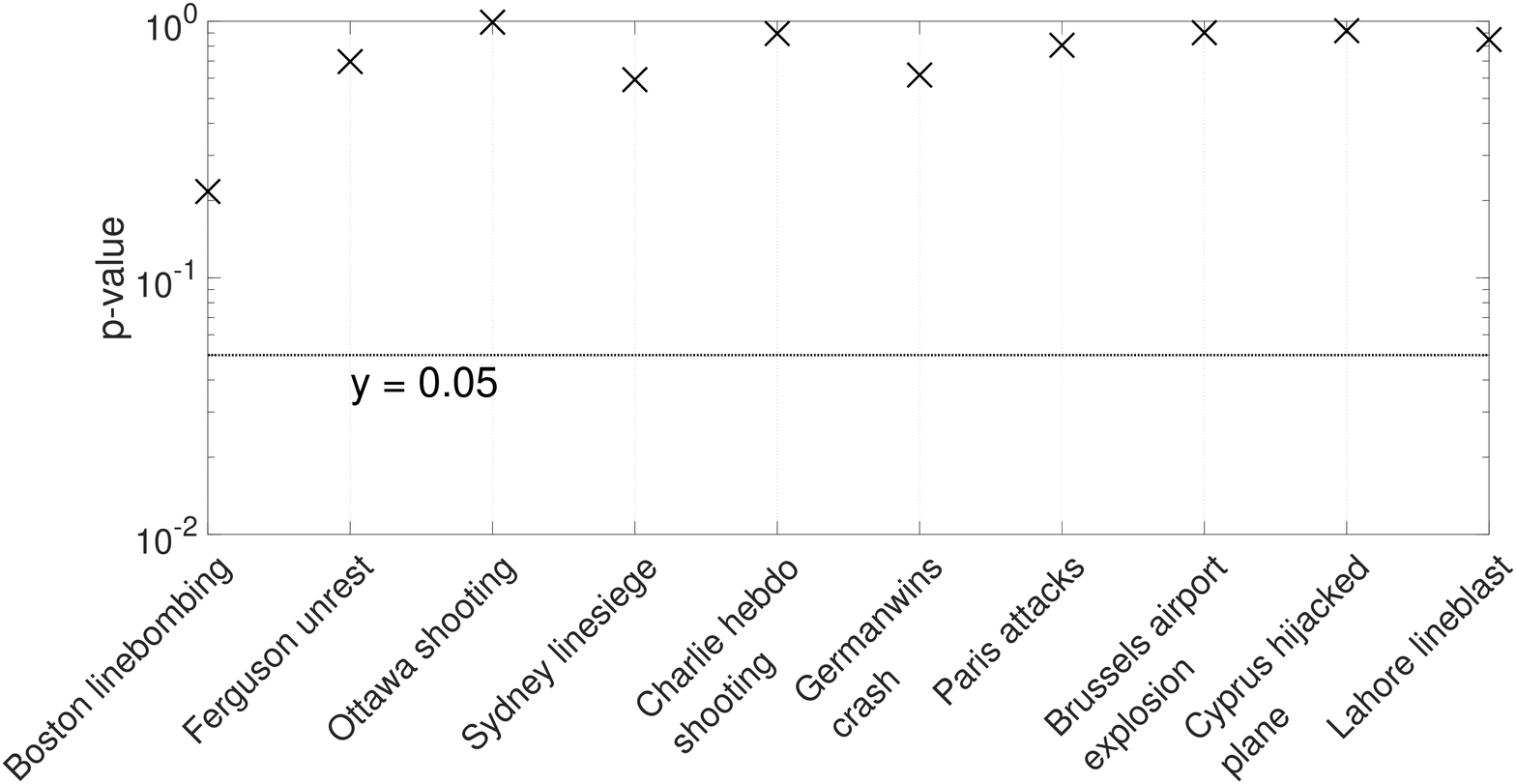}
  \caption{Local tweets (tweets with a location label that is close to where the event occurred) only.}
  \label{figure_p_value_2012_2016_local}
\end{subfigure}
    
\caption{\(P\)-values of the significance test for \texorpdfstring{\(D_{1}-D_{30}\)}{Lg}. A \(p\)-value of less than 0.05 indicates that the power-law hypothesis should be rejected.}
\label{figure_p_value_2012_2016}
\end{figure}

Fig.~\ref{figure_p_value_2012_2016} shows the \(p\)-values for the time series of (1) the tweets with a location label (Fig.~\ref{figure_p_value_2012_2016_geo}), and (2) the tweets that are close to where the 10 short-term and unplanned outbursts have occurred (Fig.~\ref{figure_p_value_2012_2016_local}). We can see that in the first case, 24 out of 30 time series pass the significance test, while in the second case, all the 10 time series are with a \(p\)-value larger than 0.05. This indicates that when an event occurs, there is high probability that a power-law distribution can be detected in the geo-tagged tweet stream from the surrounding areas.

\subsection{Power-law Distribution in \texorpdfstring{\(D_{31}\)}{Lg}}
We continue the test of a power-law distribution for the dataset \(D_{31}\) of over 500 events to further verify the above finding. Note that we only examine the tweets associated with an event according to the provided ground truth. In addition, although most of the tweets in \(D_{31}\) do not have a location label, the majority of the events are regional, and hence it is likely that most of the tweets are posted close to where the events have occurred.

\begin{figure}[t!]
\centering
\includegraphics[width=.9\columnwidth]{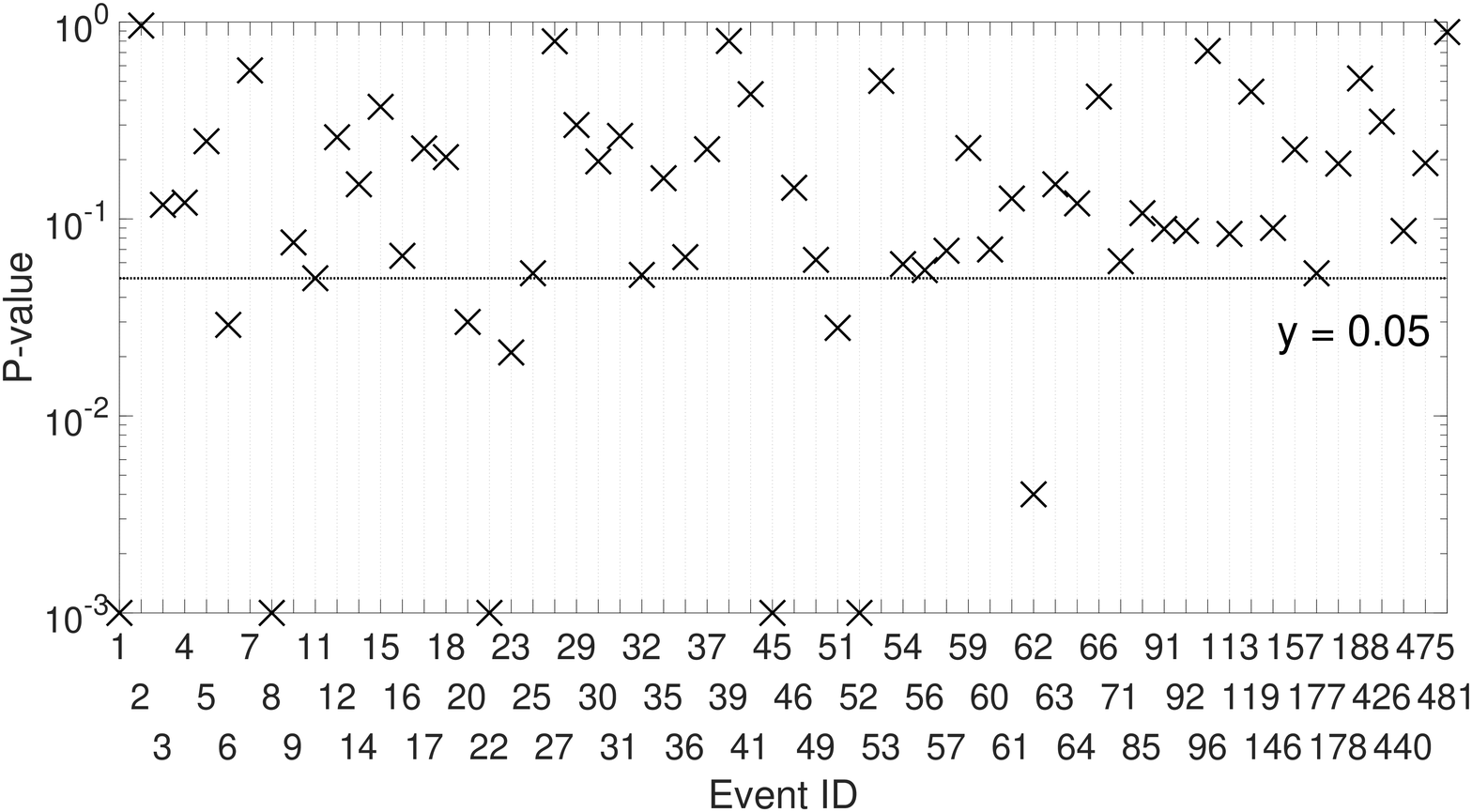}
\caption{\(P\)-values of the significance test for \texorpdfstring{\(D_{31}\)}{Lg}. For Events 1, 8, 22, 45 and 52, the original \(p\)-values of 0.0 are replaced with 0.001, in order for them to be shown in the figure.}
\label{figure_p_value_2012}
\end{figure}

As can be seen from Fig.~\ref{figure_p_value_2012}, 52 out of the 62 time series pass the significance test. The above statistics in Figs.~\ref{figure_p_value_2012_2016}-\ref{figure_p_value_2012} indicate that \textbf{\emph{when an event happens, it is likely that the time series corresponding to the geo-tagged tweets from surrounding areas follows a power-law distribution, and hence exhibits self-similarity.}}

Up till now, we have only considered tweets of certain events. However, can a power-law distribution be observed as well when no event occurs (\ie false positives)? In order to answer this question, we randomly extract 100 two-hour intervals from all tweets with a location label, remove those tweets that are associated with any of the 506 events, and check whether a power-law distribution can be detected within the 100 time series corresponding to the remaining tweets, where each time series counts the number of tweets posted every minute during the two-hour interval. The result shows that only 21 of the time series follow a power-law distribution---the percentage is much lower than when an event occurs.

Finally, we consider the overall case where all tweets are mixed together, no matter if they are associated with any event or not, and check whether a power-law distribution can be detected to further examine the probability of false positives. Specifically, we extract 1000 two-hour intervals (potentially with overlap) randomly from all tweets with a location label, and then validate the existence of a power-law distribution in the generated time series. In this case, 25.0\% of them pass the significance test, which is also obviously lower than the percentage when an event occurs.

\subsection{Power-law Distribution in \texorpdfstring{\(D_{32}\ \&\ D_{33}\)}{Lg}}
In order to verify the last observation in the above subsection, \ie the overall case, we further test datasets \(D_{32}\ \&\ D_{33}\), both of which include all geo-tagged tweets from a certain area, not just specific to any events. We take the same approach by randomly selecting 1000 two-hour windows from each dataset, and run the significance test on the corresponding time series. The results are listed in Table~\ref{table_melb_ny_p_values}. We believe the high percentage for \(D_{32}\) (New York) at Level 0 is because there are significantly more tweets in this dataset, which contains too much noise and leads to false positives. In fact, as can be seen in Table~\ref{table_melb_ny_p_values}, if we zoom into a sub-region of New York, and generate the time series by counting the tweets only from there, the percentage decreases quickly.

\begin{table}[t!]
\caption[caption]{Percentage of test windows where a power-law distribution can be observed at different spatial scales in the datasets of \texorpdfstring{\(D_{32}-D_{33}\)}{Lg}.}
\begin{center}
\begin{tabular}{|c|c|c|}
\hline
 & \textbf{\(D_{32}\): New York} & \textbf{\(D_{33}\): Melbourne} \\
\hline
Level 0: overall & 48.1 & 5.2 \\
\hline
Level 1: sub-region 1-1 & 17.1 & 0.2 \\
\hline
Level 1: sub-region 1-2 & 10.7 & 0.6 \\
\hline
Level 1: sub-region 1-3 & 12.2 & 0.0 \\
\hline
Level 1: sub-region 1-4 & 14.0 & 0.0 \\
\hline
Level 2: sub-region 2-1 & 11.4 & 0.1 \\
\hline
Level 2: sub-region 2-2 & 7.0 & 0.2 \\
\hline
Level 2: sub-region 2-3 & 17.1 & 0.0 \\
\hline
Level 2: sub-region 2-4 & 11.8 & 0.0 \\
\hline
Level 3: sub-region 3-1 & 10.8 & 0.1 \\
\hline
Level 3: sub-region 3-2 & 9.9 & 0.0 \\
\hline
Level 3: sub-region 3-3 & 21.2 & 0.0 \\
\hline
Level 3: sub-region 3-4 & 10.6 & 0.0 \\
\hline
\end{tabular}
\label{table_melb_ny_p_values}
\end{center}
\end{table}

The above experimental results suggest that \textbf{\emph{for a collected set of tweets, if a considerable portion of them are about a certain event, then a power-law distribution is likely to be observed in the corresponding time series}}. Therefore, we propose to use the existence of a power-law distribution to help detect or verify events from geo-tagged tweet streams. In the next section, we test this idea by building a simple event detection algorithm that ignores the content of a tweet, but only counts the number of tweets posted during a short time period at different geographic scales, and checks whether it follows a power distribution.

\section{Application in Event Detection}\label{sec:application}
\update{
This section aims to apply the previous finding of the correlation between the occurrence of an event and a power-law distribution in tweet streams for event detection. We first propose an algorithm \textit{Power-law basic} and show that by checking power-law distributions alone, it can achieve comparable results to more complex algorithms that use semantic analysis in addition to spatial clustering, \eg Geoburst~\cite{zhang_geoburst:_2016}, a popular state-of-the-art event detection algorithm. Then we integrate semantic analysis with power-law verification, and show that this improved version, \textit{Power-law advanced}, can achieve significantly better performance.
}

\subsection{\update{Power-law Basic: Power-law based Multi-scale Spatial Event Detection}}
We start with a brief problem definition of event detection from tweet streams. For a certain region \(R\), given a stream of tweets \(T = \{t_{1}, t_{2}, ..., t_{n}\}\) and a query window \(W = \{t_{n-m+1}, t_{n-m+2}, ..., t_{n}\}\) (\(m\) is the number of tweets in \(W\)) that represents currently observed tweets, the aim is to identify a set of tweets \(T_{i} \subseteq W\) that are associated with an event as close to where and when the event occurs as possible.

To solve the above problem, we propose to create a Quad-tree (\(QT\)) for each \(W\), the root of which represents the whole region \(R\). If \(m\) exceeds the predefined threshold \(m_{s}\), \(QT\) divides \(R\) into four equally sized sub-regions, and the process continues until the number of tweets in each leaf node is not larger than the threshold, or the depth of \(QT\) reaches the maximum value, \ie the size of a sub-region has to be larger than a certain value. Once the Quad-tree is built, the detection will be run at all levels, which mitigates the impact of the arbitrary division of space.

As shown in Algorithm~\ref{algo:detection}, we check for the existence of a power-law distribution in each node of \(QT\): for a node \(N\), (1) collect tweets from all children nodes recursively (note that once a node is divided, it does not hold any tweet itself, as all its tweets are moved to one of the four child nodes); (2) divide the query window into multiple time intervals of \(d\) seconds, and count the number of tweets posted in each interval to generate the time series \(S\) (here \(d\) does not need to be 60 as in our previous experiments, \eg a time series of tweets posted every 30 seconds should still follow a power-law distribution); (3) fit a power-law model to \(S\); (4) run the significance test and calculate the \(p\)-value; (5) reject the power-law hypothesis if the \(p\)-value is less than 0.05, otherwise create a new event with all the tweets and append it to the final result; (6) repeat (1)-(5) for each child node at the lower levels, so that an event can be detected as close to where it happens as possible.

\subsubsection{Delay in the Validation of a Power-law Distribution}
An important question is: how many data points need to be observed (\(n_{min}\)), \ie the minimum length of the time series, or the delay, to verify a power-law distribution? The answer impacts two important parameters in the above algorithm: the length of the query window, \(l\) (in seconds; \(l\) is different from \(m\), which is the number of tweets in a query window), and the length of the time interval, \(d\) (in seconds), since \(n_{min} = l/d\). For example, if \(n_{min} = 100\) and \(l = 600\), then in order to obtain 100 data points from each query window, the algorithm divides the window into 100 intervals, \ie counts the number of tweets posted every 6 seconds.

Our experiments on datasets \(D_{1}-D_{31}\) suggest that when \(l\) and \(d\) are chosen properly, so that the majority of the elements in the time series are above zero, then the power-law distribution can be verified using the first 60 data points, \ie \(n_{min} = 60\). Normally, a larger value of \(n_{min}\) contributes to a lower false positive rate, but too large a value causes few events to be found, which decreases the precision. In the following experiments, we set \(60 \leq n_{min} \leq 300\ and\ 1200 \leq l \leq 3600\). A more detailed sensitivity analysis is given in the next subsection.

\begin{algorithm}[t!]
\LinesNumbered
\SetKwInOut{Input}{\small Input}
\SetKwInOut{Output}{\small Output}
\Input{Geo-tagged tweets in the query window, \(W\); \\ 
Maximum depth of the Quad-tree (QT), \(D\); \\
Threshold for splitting a node in QT, \(m_{s}\); \\
Length of the query window, \(l\); \\
Time interval to count tweets, \(d\) (seconds)}
\Output{Event list, \(E\)}
\BlankLine

\SetKwFunction{algo}{algo}\SetKwFunction{f}{f}
\SetKwProg{buildqt}{Phase 1: Build Quad-tree}{}{}
\buildqt{}{
    Create an empty Quad-tree \(QT\)\;
    \For{tweet \(t\) in \(W\)}{
        \If{child nodes != NULL}{
            Insert \(t\) into one of the child nodes based on \(t\)'s coordinates\;
        }
        \ElseIf{the number of tweets in the current node \(\geq m_{s}\) \&\& \(QT\)'s depth \(< D\)}{
            Split the current node into four nodes\;
            Move all tweets including \(t\) into one of the four child nodes according to the coordinates\;
        }
        \Else{
            Insert \(t\) into the current node\;
        }
    }
}{}

\SetKwProg{entdt}{Phase 2: Multi-scale spatial event detection}{}{}
\entdt{\f{\(N, l\)}}{
    \For{node \(N\) in \(QT\)}{
        Collect tweets from all children nodes recursively\;
        Generate the time series \(S\): divide \(l\) into multiple intervals of \(d\) seconds and count the number of tweets posted during each interval\;
        Fit a power-law model to \(S\)\;
        Run the significance test, and calculate the \(p\)-value for the fitted model\;
        \If{\(p\)-value \(<\) 0.05}{
            Reject the power-law hypothesis\;
        }
        \Else {
            Create a new event, append all tweets from children nodes to it, and insert it to \(E\)\;
        }
        \tcp{Detect events at lower levels recursively}
        \For{\(N^{\prime}\) in child nodes}{
        	\(E.add(f(N^{\prime}, l))\)
        }
    }
}

\Return{\(E\)}
\caption{\update{Power-law basic: power-law based multi-scale spatial event detection}}\label{algo:detection}
\end{algorithm}

\subsection{Experimental Verification}
In order to demonstrate the performance of Algorithm~\ref{algo:detection}, we have tested it against Geoburst on three datasets (we did not choose the improved versions of Geoburst+~\cite{zhang_geoburst+:_2018} and TrioVecEvent~\cite{zhang_triovecevent:_2017} because they use supervised approaches, while both Geoburst and our method use unsupervised approaches. In addition, the purpose here is just a proof of concept that power-law verification can be used for event detection):

\begin{itemize}
    \item All geo-tagged tweets from Melbourne in Jan 2017, with a size of 23.3K;
    \item All geo-tagged tweets from Los Angeles between 9 February and 22 February 2019, with a size of 13.2K;
    \item All geo-tagged tweets from Sydney between 12 February and 5 April 2019, with a size of 28.4K.
\end{itemize}

These three datasets have different levels of event density: Melbourne \(>\) LA \(>\) Sydney, and we intend to check the performance of our method in all these settings. Specifically, the Melbourne dataset contains the event of ``Melbourne car attack"~\cite{noauthor_january_2019}, which is the type of event that we are most interested in detecting.

The reason why we do not use \(D_{31}\), although it provides the ground truth, is the lack of accurate location information and publication time: the location is normally a city/town name, and the publication time only has a precision of a minute. As a result, in order to generate a time series with a minimum length of 60, it is necessary to collect tweets for 60 minutes, and since the Quad-tree cannot divide the root node due to the missing coordinates, the algorithm always needs to check the power-law distribution at level 0 against hundreds of thousands of tweets from worldwide, which contains too much noise.

\subsubsection{Quantitative Analysis}
Depending on the density of the data, the parameters are chosen as follows to ensure that there are an appropriate number of tweets in each query window, and sufficient elements in the generated time series are above zero: (1) for the dataset collected from Melbourne, each query window is set to 30 minutes, \ie \(l = 1800\), \(n_{min}\) is set to 80, and \(m_{s} = 15\); (2) for the dataset collected from LA, \(l = 1200\), \(n_{min} = 150\), and \(m_{s} = 50\); (3) for the dataset collected from Sydney, \(l = 3600\), \(n_{min} = 100\), and \(m_{s} = 50\). To make the results comparable, we set the query window to be of the same length for Geoburst, and all other parameters take the default values in the code shared by the author. For each of the three datasets, we run both algorithms on consecutive query windows covering the whole period. For example, the Melbourne dataset lasts 31 days, so there are 31\(\times\)48 = 1488 query windows.

\begin{figure}[t!]
\centering

\begin{subfigure}{\columnwidth}
  \centering
  \includegraphics[width=\columnwidth]{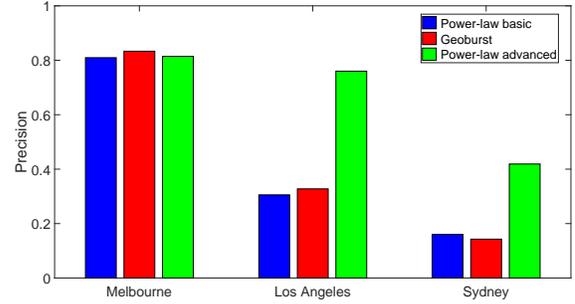}
  \caption{Precision.}
  \label{figure_precision}
\end{subfigure}

\begin{subfigure}{\columnwidth}
  \centering
  \includegraphics[width=\columnwidth]{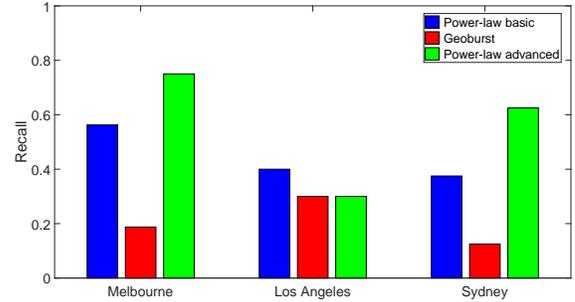}
  \caption{Recall.}
  \label{figure_recall}
\end{subfigure}

\caption{Performance comparison of the three event detection algorithms.}
\label{figure_performance_comparison}
\end{figure}

Fig.~\ref{figure_performance_comparison} presents the performance comparison between the two algorithms, which suggests that even though \textit{Power-law basic} does not check the content of each tweet, it achieves comparable performance with Geoburst, in terms of both precision and recall. Note that since the ground truth of the three datasets are not given, it is difficult to calculate the true recall. Therefore, we adopt a similar approach as in~\cite{zhang_geoburst+:_2018,zhang_triovecevent:_2017} and calculate the \(pseudo\ recall = N_{true}/N_{total}\), where \(N_{true}\) is the number of true events detected by a method, and \(N_{total}\) is the number of true events detected by all methods, plus the events hand-picked by us that occurred during the query periods within the chosen cities, including festivals, sport games, natural disasters, etc. Note also that the validity of each event is checked manually.

However, we are not claiming that it is sufficient to detect events just by checking the existence of a power-law distribution, and we further improve the algorithm in Section~\ref{sec:discussion}.

\begin{figure}[t!]
\centering
\includegraphics[width=.9\columnwidth]{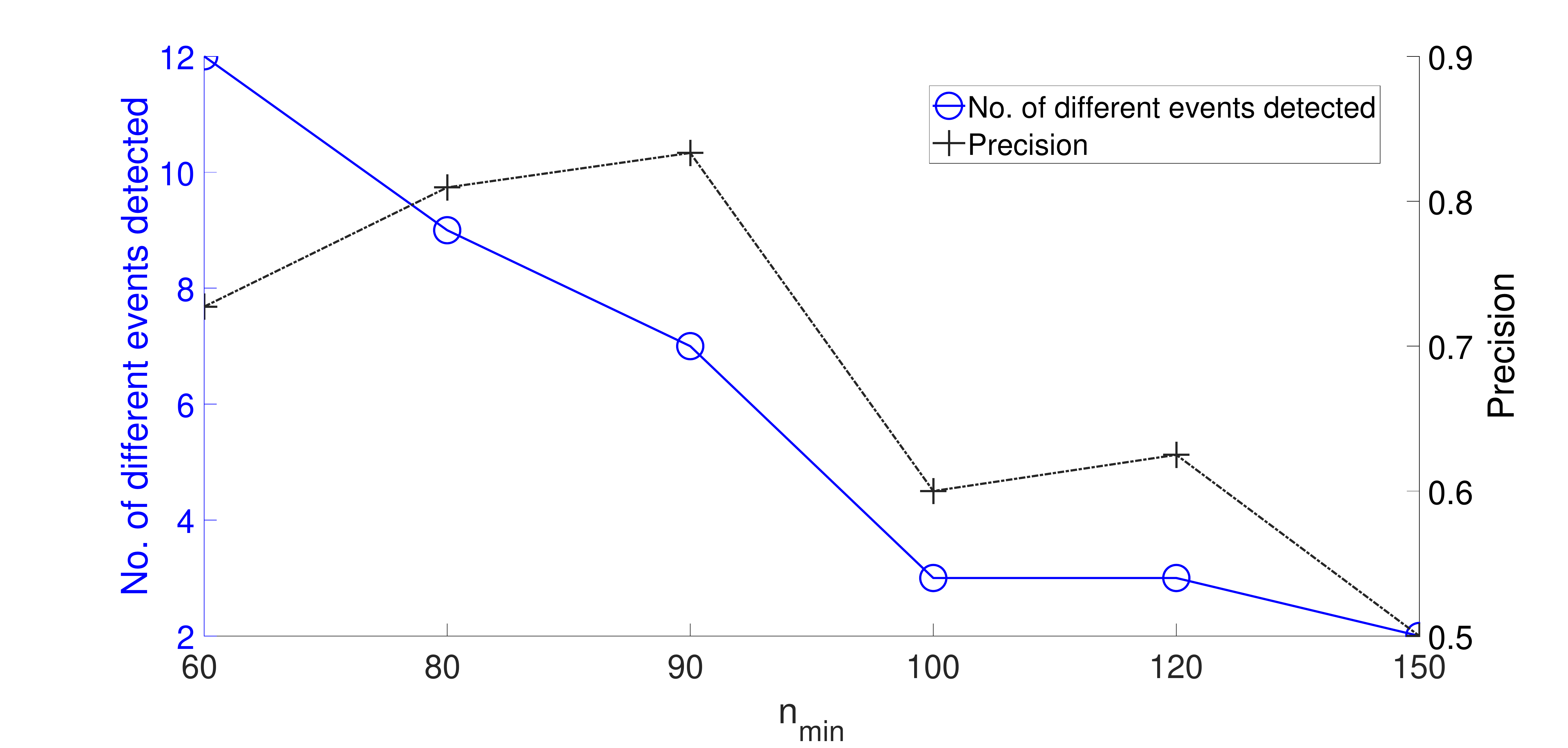}
\caption{Sensitivity analysis on \(n_{min}\) for the Melbourne dataset. The length of the query window \(l\) is set to 30 minutes, so \(n_{min} = 60, 80, 90, 100, 120, 150\) corresponds to counting the number of tweets every 30, 22.5, 20, 18, 15, 12 seconds.} 
\label{figure_n_min}
\end{figure}

\textbf{Sensitivity analysis on \(n_{min}\).} Fig.~\ref{figure_n_min} shows how \(n_{min}\) impacts the number of different events detected and the precision for the dataset of Melbourne, when \(l\) is set to 30 minutes and \(m_{s} = 15\). As \(n_{min}\) first increases, both the reported events and false positives decrease, and the false positive count decreases faster, so the precision improves. However, as \(n_{min}\) gets too large, \ie counting the number of tweets too frequently, too many elements of the generated time series become zero, causing too few events to be detected, and the precision drops.

\begin{figure}[t!]
\centering
\includegraphics[width=.9\columnwidth]{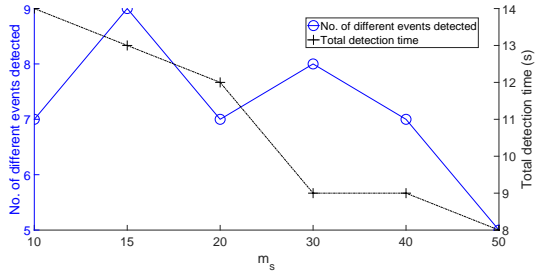}
\caption{Sensitivity analysis on \(m_{s}\) for the Melbourne dataset. The length of the query window \(l\) is set to 1800 (30 minutes), and \(n_{min}\) is set to 80.} 
\label{figure_m_s}
\end{figure}

\textbf{Sensitivity analysis on \(m_{s}\).} We further analyse the impact of \(m_{s}\). Specifically, Fig.~\ref{figure_m_s} shows for the dataset of Melbourne, when \(l=1800\) and \(n_{min}=80\), how the number of different events detected and the total detection time change with \(m_{s}\). A small \(m_{s}\) value means a larger depth of the Quad-tree, and since the detection is running at each node, the total detection time will be longer, but meanwhile more events are likely to be found.

\update{
\subsection{\textit{Power-law Advanced}: Combining Semantic Analysis with Power-law Verification}\label{sec:discussion}
}
In order to further improve the performance of the proposed method, we investigate how semantic information can be incorporated to the event detection algorithm. 

A common class of existing methods that use semantic information is clustering based approaches, where the first step is to cluster posts/tweets according to their semantic, spatial, temporal, frequency information, etc., and then generate a list of event candidates. Once the candidates are found, the second step is to remove non-event clusters among them. Our finding in this paper suggests that checking the existence of a power-law distribution can be used in the second step to test whether a cluster of tweets are about a real event or not.

In order to demonstrate the feasibility of the above approach, we design an algorithm \textit{Power-law Advanced} that combines fastText (the latest word embedding tool developed by Facebook)~\cite{bojanowski_enriching_2016}, BIRCH (Balanced Iterative Reducing and Clustering using Hierarchies)~\cite{zhang_birch:_1996}, and power-law verification, where fastText is for embedding the tweets so that semantically similar tweets would also end up close in the vector space, and BIRCH is for clustering the generated vectors. These two methods are chosen for demonstration purposes only, and they can be replaced by other alternatives.

In addition, sliding windows are used: the algorithm keeps the latest \(N_{SW}\) query windows, performs event detection, and discards the oldest window while collecting new tweets. In the following experiment, \(N_{SW}\) is set to 6, and the size of a query window is set to 30 minutes. 

\begin{figure}[t!]
\centering
\includegraphics[width=.9\columnwidth]{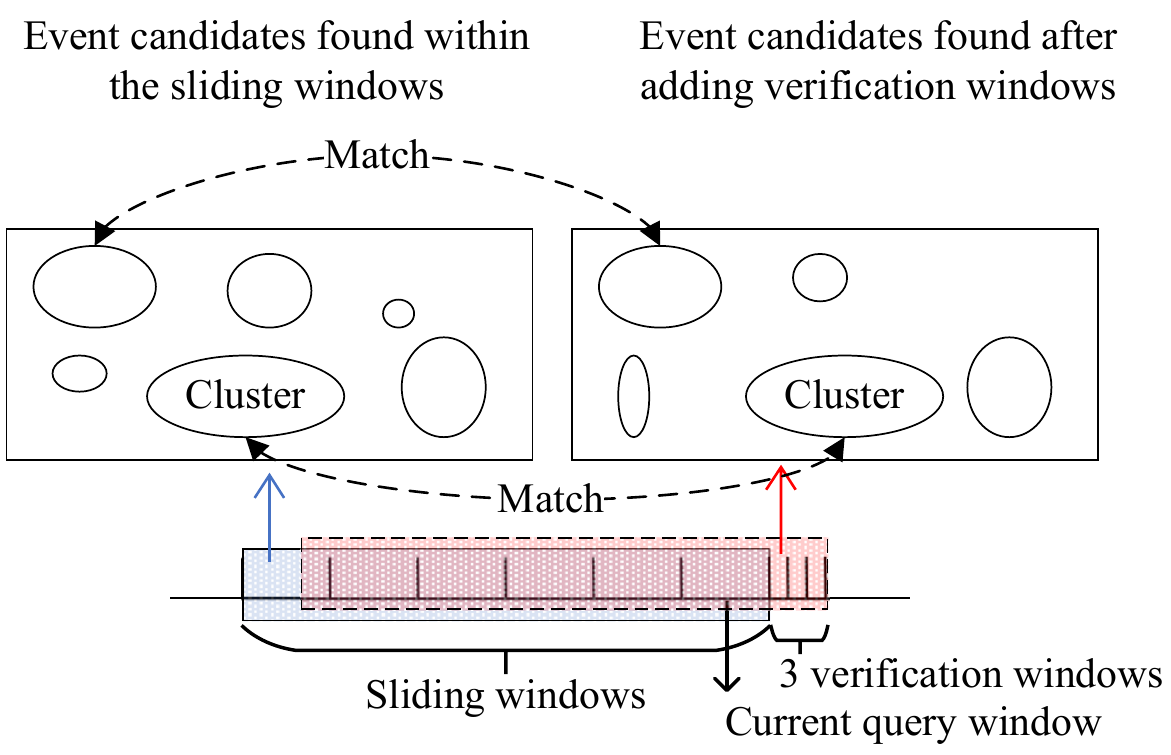}
\caption{An illustration of \textit{Power-law Advanced}.} 
\label{figure_illustration}
\end{figure}

Specifically, the algorithm (Algorithm~\ref{algo:detection_advanced}: \textit{Power-law Advanced}) works as described below (also see Fig.~\ref{figure_illustration} for an illustration):
\begin{itemize}
    \item Embedding. The same NLP tool~\cite{ritter_twitter_2019} as mentioned in ~\cite{zhang_geoburst:_2016} is used to extract entities and noun phrases from the tweets. These generated keywords are then embedded with the fastText algorithm, and each tweet is represented by the average value of the vectors from all its keywords. A pre-trained fastText model is used in our experiment, and it is re-trained incrementally with the new tweets~\cite{qinluo_library_2019}. The re-training is done in parallel, and hence does not delay the detection. Note that the spatial and temporal information is not included in the embedding since the Quad-tree and sliding windows ensure the similarity in terms of space and time.
    \item Clustering. Once the tweets are embedded into vectors, we use the BIRCH algorithm to cluster the vectors. The most important parameter in BIRCH is the threshold of the cluster radius. In our experiment we do not directly set a fixed value. Instead, we start with a value close to zero, and increase it by a small step size until either (1) less than 5\% of all items are in small clusters, \ie clusters with a size less than 10, or (2) over half of the items are in the largest cluster, whichever occurs first.
    \item Power-law detection. The third step is to detect any power-law distribution within each cluster. Note that a Quad-tree is still built and maintained, and the detection is run at all levels of the Quad-tree to mitigate the impact of the arbitrary division of space.
    \item Verification. If any event candidate is found in the last step, we further collect tweets from the verification window which is set to 5 minutes in our experiment, and repeat the above three steps. The only difference is that the keywords are no longer used, and the original text of each tweet is directly embedded---the rationale is to ensure that both the keywords and texts are semantically close within a cluster. Each event candidate from the last step is then checked against each cluster found in this step. If any two of them share more than half of the tweets, they are considered as a match. If no match is found for a candidate, it will be removed. The verification process is done three times, and an event candidate has to pass all three of them.
    \item Final clean-up. To further decrease the false positive rate, the last step extracts the top \(X(=10)\) hashtags and mentions for each cluster, and if more than half of the tweets contain any of these hashtags or mentions, the cluster is finally considered as an event.
\end{itemize}

\begin{algorithm}[t!]
\update{
\LinesNumbered
\SetKwInOut{Input}{\small Input}
\SetKwInOut{Output}{\small Output}
\Input{same as \textit{Power-law basic}}
\Output{Event list, \(E\)}
\BlankLine

Extract entities and noun phrases using the NLP tool~\cite{ritter_twitter_2019} for each tweet\;
Call fastText to embed the extracted keywords\;
Cluster the generated vectors using BIRCH\;
\For{Cluster \(c\) found in the last step}{
    \(E.add(\textit{Power-law basic}(\cdot))\);
}
\For{\(i=0;\ i<3\ \&\&\ E\) is not NULL}{
    Call fastText to directly embed the text of each tweet\;
    Cluster the generated vectors using BIRCH\;
    \For{Cluster \(c^{\prime}\) found in the last step}{
        \(E^{\prime}.add(\textit{Power-law basic}(\cdot))\)\;
    }
    \For{Remaining event candidate \(e \in E\) }{
        \If{there is no match in \(E^{\prime}\)}{
            Remove \(e\);\
        }
    }
}
\For{Remaining event candidate \(e \in E\) }{
    \(K \leftarrow\) Extract the top \(X(=10)\) hashtags and mentions\;
    \If{More than half of the tweets in \(e\) does not contain any element in \(K\)}{
        Remove \(e\);\
    }
}

\Return{\(E\)}
\caption{Power-law advanced: integrating power-law verification with semantic analysis}\label{algo:detection_advanced}
}
\end{algorithm}

We test the above algorithm on the same three datasets, and as can be seen in Fig.~\ref{figure_performance_comparison}, this semantic analysis enhanced power-law verification method increases both the precision and the recall in most cases.

In summary, this section has demonstrated that the naive algorithm of checking the existence of a power-law distribution can achieve comparable results against more advanced event detection methods, and its performance can be significantly improved by integrating with semantic analysis.

\section{Related Work}\label{sec:related_work}
This section briefly reviews the previous work on event detection from social media. Specifically, we take a similar approach as~\cite{panagiotou_detecting_2016} and summarise two types of algorithms: clustering based and anomaly based. In addition, multiscale event detection is also considered.

\subsection{Clustering based Event Detection}
This type of detection method takes into consideration all or a subset of temporal, spatial, semantic, frequency and user information to cluster the tweets~\cite{becker_beyond_2011,li_tedas:_2012,ozdikis_semantic_2012,abdelhaq_eventweet:_2013,fuchs_tracing_2013,walther_geo-spatial_2013,xie_topicsketch:_2016,zhang_geoburst:_2016,zhang_triovecevent:_2017,zhang_geoburst+:_2018,wei_detecting_2018}. However, since the generated clusters may correspond to non-events, normally another step is taken to eliminate false positives, \eg by ranking the candidates based on certain criteria, or training a classifier to decide whether a candidate is a real event.

For example, for each pair of tweets, Geoburst~\cite{zhang_geoburst:_2016} uses the Epanechnikov kernel to calculate their geographical impact, and uses the random-walk-with-restart algorithm to to obtain the semantic impact. In this way, they identify a list of clusters of geographically close and semantically coherent tweets, \ie event candidates. Finally, historical activities are used to rank these candidates and the top \(K\) events are returned. As the improved versions, (1) Geoburst+~\cite{zhang_geoburst+:_2018} replaces the ranking algorithm in Geoburst with a candidate classification module, which learns the latent embeddings of tweets and keywords. Then together with the activity timeline, the module extracts spatial unusualness and temporal burstiness to characterise each candidate event; (2) TrioVecEvent~\cite{zhang_triovecevent:_2017} learns multimodal embeddings of the location, time and text, and then performs online clustering using a Bayesian mixture model.

\subsection{Anomaly based Event Detection}
This type of method aims to identify abnormal observations in word usage, spatial activity, sentiment levels, etc. For example, Valkanas and Gunopulos~\cite{valkanas_event_2013,valkanas_how_2013} use sentiment analysis for event detection, which is based on the idea that the sentiment level fluctuates as people respond to an event to express their opinions. Another example is to detect peaks in Twitter hashtags using a Discrete Wavelet Transformation~\cite{cordeiro_twitter_2011}, since these peaks are likely to correspond to real-world events. Specifically, only the hashtags are used, and all the remaining tweet text is discarded. In addition, Vavliakis \etal~\cite{vavliakis_event_2012} propose Latent Dirichlet Allocation based event detection for MediaEval Benchmark 2012~\cite{noauthor_mediaeval_nodate}, where the dataset contains 167,000 images from Flickr. They detect peaks in the number of photos assigned to each topic, and identify an event for a topic if it receives an unexpectedly high number of photos.

\subsection{Multiscale Event Detection}
Running event detection on a fixed spatial resolution may not help in finding events at different scales. For example, using low resolution spatial data might only capture events occurring on the state or the country level, while high resolution data can help detect events at community or city scales. Therefore, another stream of work intends to detect events at different space resolutions, to better adapt to the unpredictability of real-life events~\cite{dong_multiscale_2015,capdevila_scaling_2016,visheratin_multiscale_2018}. For example, Dong \etal~\cite{dong_multiscale_2015} explore the properties of the wavelet transform for the detection of events at different spatio-temporal scales. In addition, Visheratin \etal~\cite{visheratin_multiscale_2018} build a convolutional quad-tree, which instead of dividing a region into four sub-regions of equal size, uses a convolutional neural network to decide a more appropriate division.

\section{Conclusions and Future Work}\label{sec:conclusions}
In this paper, we have (1) verified in more than 30 datasets the existence of self-similarity in the time series of the number of geo-tagged tweets posted within a short time interval from a certain region; (2) demonstrated that a power-law distribution is much more likely to be observed when an event occurs in tweet streams; (3)\update{ proposed two event detection algorithms: \textit{Power-law basic} and \textit{Power-law advanced}. \textit{Power-law basic} is based on the validation of power-law distributions at multi-spatial scales, without checking the content of each tweet, or using any information other than the geo-location. Experimental results on multiple datasets show that it can achieve comparable performance with Geoburst, a widely cited event detection algorithm. \textit{Power-law advanced} improves the algorithm by incorporating semantic analysis via word embedding, and our results demonstrate that it can significantly increase both the precision and the recall.}

For future work, we will further study the self-similar patterns in tweet streams. Our current result explains why when an event occurs the corresponding time series shows self-similarity---it follows a power-law distribution. \update{However, we have not examined why the tweet count time series still exhibits self-similarity when there is no event.}

\update{In addition, as a separate direction, we will explore other potential ways for embedding and clustering to futher improve the performance of the event detection algorithm. Specifically, in terms of embedding, we are considering (1) dispensing with the Quad-tree and directly embed the location information; (2) representing a tweet using other methods rather than the average value of the vectors for each word that it contains.}

\section*{Acknowledgement}
This research is funded in part by the Defence Science and Technology Group, Edinburgh, South Australia, under contract MyIP:7293.

\bibliographystyle{IEEEtran}
\bibliography{./bibliography/references}

\end{document}